\documentclass[aps,prl,twocolumn,showpacs,preprintnumbers,amsmath,superscriptaddress]{revtex4}
\usepackage{bm}
\usepackage{amsmath}
\usepackage[dvips]{graphicx}

\begin{document}
\bibliographystyle{apsrev}

\title{Correlated electron current and temperature dependence of 
the conductance of a quantum point contact}

\author{C. Sloggett}
\email{clares@phys.unsw.edu.au}
\affiliation{School of
Physics, University of New South Wales, Sydney 2052, Australia}

\author{A. I. Milstein}
\email{A.I.Milstein@inp.nsk.su} 
\affiliation{Budker
Institute of Nuclear Physics, 630090 Novosibirsk, Russia}
\author{O. P. Sushkov}

\email{sushkov@phys.unsw.edu.au}
\affiliation{School of
Physics, University of New South Wales, Sydney 2052, Australia}

\date{\today}
\begin{abstract}
We investigate finite temperature corrections to the Landauer formula
due to electron-electron interaction within the quantum point contact.
When the Fermi level is close to the barrier height, the interaction is
strongly enhanced due to semiclassical slowing of the electrons.  To
describe electron transport we formulate and solve a nonlocal kinetic
equation for the density matrix of electrons.  The correction to
the conductance $G$ is negative and strongly enhanced in the region
$0.5\frac{2e^2}{h} \le G \le 1.0\frac{2e^2}{h}$.  Our results for
conductance agree with the so-called ``0.7 structure'' observed in
experiments.
\end{abstract}
\pacs{
73.23.-b 
72.10.-d 
73.21.Hb 
73.63.Rt 
}
\maketitle

{\it Introduction}.
The conductance of a quantum point contact (QPC) - a 1D constriction in a 2D
electron gas - has been known to be quantized in units of $G_0=2 e^2 /
h$ since 1988 \cite{wharam88,vanwees88}. The observed conductance plateaus 
can be easily understood in the single-electron picture \cite{land,buttiker90}.
Below we write conductance in units of $G_0$.

The ``0.7 structure'' appears on the lowest conductance step
as a narrow extra plateau at $G \approx 0.7$. The structure was first observed 
by  Thomas {\it et al.} in 1996\cite{thomas96} and has been the subject of 
numerous experiments since 
\cite{thomas98, kristensen00,reilly01,cronenwett02,noise,danneau}. 
The position of the structure varies from 0.4 to 0.7 
depending on the device, and the structure becomes more pronounced as the
temperature is increased up to $\sim 2\textrm{K}$, where thermal smearing becomes 
significant. It is not clear from experiment whether the structure survives at 
$T=0$. In a longitudinal magnetic field, which breaks the spin
degeneracy, the structure evolves smoothly to the 0.5 plateau that is 
expected in the single-electron picture. This demonstrates that the effect 
is in some way related to electron spin.

Observations of the 0.7 structure have created much theoretical interest.
There have been suggested explanations based on  spontaneous magnetization 
within  the contact \cite{Chuan,Calmels,Zabala,bruus01,starikov03,reilly05},
charge density waves within the contact \cite{spivak,sushkov}, the Kondo 
effect \cite{meir,cornaglia04}, and even on  electron-phonon scattering in 
the contact \cite{seelig03}. The problem has also stimulated the development
of general scattering theory in the presence of leads \cite{molina,meden}.

In the present paper, using perturbation theory, we consider the 
finite temperature correction to the Landauer formula  for the 
conductance of a QPC. Our results are very similar to  experimental data on 
the 0.7 structure.

{\it Model}.
Near the center of the QPC the single-particle dynamics are described by the 
parabolic saddle-point potential
$U=U_0-\frac{1}{2}m\omega^2x^2+\frac{1}{2}m\omega_y^2y^2$, 
see, e.g., Ref. \cite{buttiker90}. 
Here $m$ is the effective mass, $U_0$ is an energy that can
be controlled in experiment using the gate voltage, and $\omega_y$
and $\omega$ are parameters with typical values in experiment of $\omega_y/\omega \sim 3$, $\omega_y \sim 4\textrm{meV}$.
Throughout the paper we set $\hbar=k_B=1$. 
The dynamics in the x-direction are described by
different channels corresponding to quantization in the y-direction.
The lowest, $n=0$,  channel has longitudinal potential
$-\frac{1}{2}m\omega^2x^2$, and hence the transmission probability is \cite{BM}
\begin{equation}
\label{t}
t_{\epsilon}=\left(1+e^{-2\pi\epsilon}\right)^{-1} \ ,
\end{equation}
where $\epsilon=E/\omega$, $E$ is the energy.
At $\epsilon =0$ the turning point for the
$n$th channel is $x_t=\pm \sqrt{\frac{2n\omega_y}{m\omega^2}}$.
So outside the barrier, from $x_{2D}\sim 2/\sqrt{m\omega}$,
many transverse channels are occupied and hence the
electron-electron Coulomb interaction is strongly screened. The interaction is 
unscreened only on the top of the barrier where  the electron density is low.
Therefore the effective electron-electron interaction in the $n=0$ channel
can be approximated as
\begin{equation}
\label{int}
\frac{e^2}{\kappa|x_1-x_2|}\to
H_{int}=\omega\pi^2g_e \delta(\xi_1)\delta(\xi_2)  \ ,
\end{equation}
where $e$ is the electron charge, $\kappa$ is  the dielectric constant,
and $\xi=\sqrt{m\omega}x$ is the dimensionless distance.
In GaAs, $\kappa\approx 13$, $m\approx0.07m_e$, and
for $\omega \sim 1\textrm{meV}$ the dimensionless coupling constant 
$g_e\sim \frac{e^2}{\pi^2\kappa}\sqrt{m/\omega}$ is about unity, $g_e\sim 1$.
Another issue related to the many-body screening is the absence of
pronounced structures  in  higher conductance steps. 
For higher steps there are always lower channels penetrating the QPC.
This leads to high electron  density in the contact and hence to screening 
of the effective interaction (2).

\begin{figure}[htb]
\includegraphics[scale=0.7]{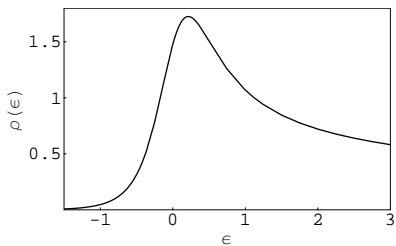}
\vspace{-10pt}
\caption{\it The probability density at 
$\xi=0$ versus the dimensionless energy $\epsilon$.}\label{rho}
\end{figure}

To represent single-particle wave functions in the $n=0$ channel we consider a 
1D wire of length $L$ with a potential  barrier in the middle of the wire. 
The details of the potential shape are unimportant apart from
the parabolic top, see, e. g., \cite{BM}.
We need the ``wire'' to normalize wave functions according to
$\int_0^L|\psi_k|^2dx =1$,
since this normalization is convenient for discussion of the nonlocal 
kinetic equation, see below.
The length $L$, which is of the order of the ballistic mean free path, 
disappears from final answers. Away from the barrier the wave function 
$\psi_{k}$ is the standard combination of incident, reflected and 
transmitted waves. Near the potential top the wave function is of the form,
see, e. g., Ref. \cite{BM}
\begin{eqnarray}
\label{psi}
&&\psi_k(x)\approx \frac{1}{\sqrt{L}}\left(\frac{mv_F^2}{2\omega}\right)^{1/4}
\varphi_k(\xi) \ ,
\nonumber\\ 
&&\varphi_{k}(\xi)=\sqrt{\frac{e^{\pi\epsilon/2}}{\cosh(\pi\epsilon)}}
D_{\nu}(\sqrt{2}\xi e^{-i\pi/4}) \ ,
\end{eqnarray}
where $v_F$ is the Fermi velocity far from the barrier,
$D_{\nu}$ is the parabolic cylinder  function, and 
$\nu=i\epsilon-\frac{1}{2}$. 
Eq.~(\ref{psi}) corresponds to the wave incident from the left, $k \ge 0$.
Due to semiclassical slowing, the probability density at the top of the 
potential,
\begin{equation}
\label{r}
\rho(\epsilon)=|\varphi_k(0)|^2=
\frac{\pi\exp(\pi\epsilon/2)}{\sqrt{2}\cosh
(\pi\epsilon)\,|\Gamma(3/4-i\epsilon/2)|^2}\, ,
\end{equation}
is peaked at $\epsilon\approx 0.2$, see Fig.~\ref{rho}. This results in
enhancement of the interaction (\ref{int}), and in the end this leads to all 
effects considered in this paper.

The current in state (\ref{psi}) is 
$j_{k}=\frac{e}{L}\frac{k}{m} t_\epsilon$,
and hence the total current obeys the Landauer formula
\begin{eqnarray}
\label{L}
&&J=2e\int\frac{Ldk}{2\pi}j_{k}n_{0k}=
\frac{2e^2}{h}t_{\mu}V \ , \nonumber\\
&&n_{0k}=n_{f}+s\frac{eV}{2\omega}n'_{f} \ .
\end{eqnarray}
Here $n_{0k}$ is the non-equilibrium occupation number under a
small applied  voltage $V$, $n_{f}=n_{f}(\epsilon)$ is the equilibrium 
Fermi-Dirac distribution,
$s=k/|k|$ shows the direction of current flow in a particular state, 
and 
$n'_{f}=-\frac{\partial n_{f}}{\partial\epsilon}\approx \delta(\epsilon-\mu)$;
$\mu$ is the chemical potential in units of $\omega$.

{\it The nonlocal kinetic equation}.
At zero temperature  the interaction (\ref{int}) in the leading order 
is described by the usual direct and exchange diagrams for the single-particle 
scattering amplitude.
The corresponding corrections renormalize the frequency $\omega$ in the single 
particle potential and  even make the effective potential slightly 
nonparabolic (flattened). 
Nevertheless this does not materially change the profile
of the transmission coefficient (\ref{t}). 
We have checked numerically that the same is true with up to
higher orders in perturbation theory.
Thus the interaction (\ref{int}) treated perturbatively
does not influence the Landauer formula and  does not materially change the
shape of the transmission coefficient at zero temperature.

Nonzero temperature requires a consideration of the details of equilibration. 
Eq.~(\ref{L}) implies nonlocal equilibration, as the scattering 
states with $s=+1$ equilibrate due to collisions  in the left lead and 
the states with $s=-1$  equilibrate due to collisions in the  right lead.
As a consequence, the many-body density matrix 
is diagonal not  in the basis of plane waves or standing waves, but
in the basis of the scattering states (\ref{psi}) 
(see also Ref. \cite{buttiker90b}).
Therefore, the occupation numbers $n_k$, the diagonal matrix elements of 
the density matrix, obey the kinetic equation 
$\frac{\partial n_k}{\partial t}=-\frac{n_k-n_{0k}}{\tau}$.
Here $\tau$ is the relaxation time in the leads. This equation is in
principle equally valid for finite bias, but in the present paper we
consider only an infinitesimal bias where $n_{0k}$ is given
by Eq.~(\ref{L}). Taking into account electron-electron interaction in the 
QPC, we obtain a collision term in the kinetic equation \cite{LP}
\begin{eqnarray}
\label{ke1}
&&\frac{\partial n_k}{\partial t}=-\frac{n_k-n_{0k}}{\tau} + \mbox{St}(n_k)\ ,
 \\
&&\mbox{St}(n_k)=
2\pi\int \frac{Ldk_1}{2\pi}\frac{Ldk_2}{2\pi}\frac{Ldk_3}{2\pi}
\left|M_{kk_1k_2k_3}\right|^2\nonumber\\
&&\times\left[n_{k_2}n_{k_3}(1-n_k)(1-n_{k_1})
-n_kn_{k_1}(1-n_{k_2})(1-n_{k_3})\right]\nonumber\\
&&\times\delta(E_k+E_{k_1}-E_{k_2}-E_{k_3}) \ .\nonumber
\end{eqnarray}
Here $M_{kk_1k_2k_3}$ is the matrix element of the interaction (\ref{int}).
This matrix element corresponds to the real transition between quantum states, 
$|k,k_1\rangle \to |k_2,k_3\rangle$. Therefore in the matrix element the 
initial and final wave functions are given by (\ref{psi}). This differs from 
the rule for the scattering amplitude,
 where initial wave functions are given by 
(\ref{psi}), while final wave functions are  
$\psi^{(-)}_k(x)=\psi^*_{-k}(x)$ 
(the Sommerfeld rule, see, e.g., Ref.\cite{QM}).

Using occupation numbers from (\ref{L}) and expanding the collision integral
up to the first power in the bias $V$,
we find the integral at small temperature, $T \ll \omega$,
\begin{eqnarray}
\label{ct3}
\mbox{St}(n_k)&=&-\frac{T^2 e VL^3}{12v_F^3}\delta(E_k-\omega\mu)\nonumber\\
&\times& \sum_{s_1s_2s_3}
\left[s+s_1-s_2-s_3\right]\left|M_{kk_1k_2k_3}\right|^2 \ .
\end{eqnarray}
All legs in the matrix element are taken at the Fermi surface, so we need only
perform summations over the directions $s_1,s_2,s_3$.
Calculating the matrix elements of interaction (\ref{int}) with the wave 
functions (\ref{psi}) we find
\begin{eqnarray}
\label{ct4}
\mbox{St}(n_k)&=&-\frac{T^2 eVL^3}{3v_F^3}s \delta(E_k-\omega\mu)\nonumber\\
&\times&\left(\left|M_{++--}\right|^2+\left|M_{+++-}\right|^2\right)\\ 
&=&-s e V \frac{\pi^4g_e^2}{6}\frac{v_F}{L}
\left(\frac{T}{\omega}\right)^2\rho^4(\mu)
\delta(E_k-\omega\mu)\ ,\nonumber
\end{eqnarray}
where $\rho(\mu)$ is given by (\ref{r}).
Eq.~(\ref{ct4}) leads to the following steady-state solution 
of the kinetic equation (\ref{ke1})
\begin{eqnarray}
\label{n1}
n_k&=&n_{f}+\frac{eV}{2}s \delta(E_k-\omega\mu)\nonumber\\
&\times&
\left\{1-\frac{\pi^4g_e^2}{3}\frac{\tau v_F}{L}\left(\frac{T}{\omega}\right)^2\rho^4(\mu)
\right\} \ .
\end{eqnarray}
Substitution of this expression into Eq.~(\ref{L}), where
$n_{0k}\to n_k$, gives an altered conductance  in units of $G_0$
\begin{equation}
\label{dt}
t_{\mu} \to {\overline t}_{\mu}=t_{\mu}
\left\{1-\frac{\pi^4g_e^2}{3}\frac{\tau v_F}{L}\left(\frac{T}{\omega}\right)^2
\rho^4(\mu)\right\} \ .
\end{equation}
Eq.~(\ref{dt}) is justified only if the temperature-dependent correction term is
small compared to unity. 
The term is due to the  current of correlated electrons.
It will be most significant within the region $0.5\frac{2e^2}{h} \le G \le 1.0\frac{2e^2}{h}$, since above $G=0.5$ electrons can flow without tunnelling. 
The length $L$ is of the order of the mean free path in the leads, 
so it is most natural to assume that the factor  
$\frac{\tau v_F}{L}$ in (\ref{dt}) is of the order of unity,
though we cannot exclude some dependence of the factor on temperature.
In the latter case the $T^2$ dependence of the correlated current
will be modified.

Note that the correction we have discussed is due to the interaction between 
electrons with opposite spins. The interaction between electrons with parallel
spins vanishes because the exchange diagram exactly cancels out the direct one
for the contact Hamiltonian (\ref{int}).
Therefore under an applied longitudinal magnetic field $B$ we should
take $\rho^4(\mu) \to \rho^2(\mu)\rho^2(\mu')$.  Here
$\mu'=\mu-(2g_s\mu_B)B/\omega$,  where $g_s$ is the gyromagnetic ratio
and $\mu_B$ is the Bohr magneton. Since 
$\rho(\mu)$ is a peaked function, see Fig.~\ref{rho}, a
magnetic field $B\sim \omega/(2g_s\mu_B) \sim 5$T effectively switches off the
interaction. 
This gives an explanation for the smooth evolution of the
0.7 structure to  the 0.5 plateau of
non-interacting electrons under a magnetic field.

\begin{figure}[htb]
\includegraphics[height=100pt,keepaspectratio=true]{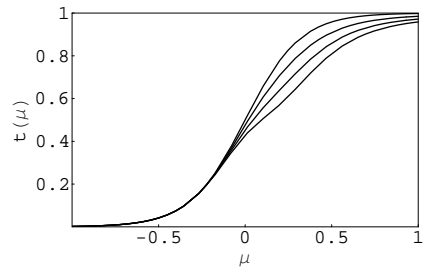}
\vspace{-10pt}
\caption{\it Conductance in units of $2e^2/h$ versus $\mu$ (chemical
potential in units of $\omega$) for different  temperatures:
the weak coupling limit,  $g_e \ll 1$. 
The uppermost curve corresponds to $g_e T =0$, while the lowest is $ g_e T \approx 0.3\textrm{K}$, assuming $\frac{\tau v_F}{L} \approx 1$.}
\label{smallg} 
\end{figure}

A set of plots of ${\overline t}_{\mu}$ for different temperatures is shown
in Fig.~\ref{smallg}.
Though the result looks quite sensible it is obtained for $g_e \ll 1$.
However, according to our estimate (\ref{int}), the constant is not
small,  $g_e\sim 1$, and hence virtual rescattering must be 
taken into account.

\begin{figure}[htb]
\includegraphics[height=50pt,keepaspectratio=true]{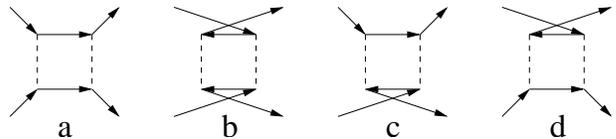}
\caption{\it The leading correction to the matrix element.} 
\label{diag} 
\end{figure}

\begin{figure}[htb]
\includegraphics[height=140pt,keepaspectratio=true]{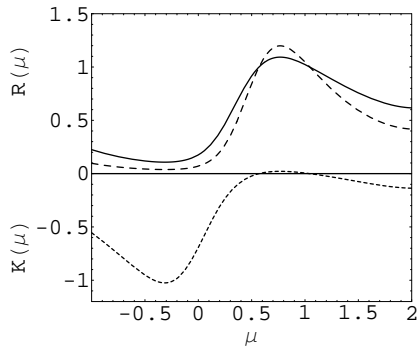}
\vspace{-10pt}
\caption{\it Dashed line: the function $K(\mu)$ for the second order 
correction.
Solid line: Brueckner correction factor $R(\mu)$ for $g_e=1$.
Long dashed line: Brueckner correction factor $R(\mu)$ for $g_e=2$.} 
\label{figK} 
\end{figure}

{\it Renormalization of the coupling constant}.
The leading correction to the Born scattering amplitude is given by the
diagrams shown in Fig.~\ref{diag}.
This correction is equivalent to renormalization of the coupling
constant, $g_e \to g_e+\delta g_e$, where $\delta g_e(\mu)=2g_e^2K(\mu)$ and
\begin{eqnarray}
\label{dg}
&&K(\mu)=\frac{1}{4}
\int\limits_{-\infty}^{\infty}\!\!\int\limits_{-\infty}^{\infty}
\rho(\epsilon_1)\,\rho(\epsilon_2)
\bigg\{\frac{\theta(\epsilon_1-\mu)\theta(\epsilon_2-\mu)}
{2\mu-\epsilon_1-\epsilon_2}\biggr.\\
&&+\biggl.
\frac{\theta(\mu-\epsilon_1)\theta(\mu-\epsilon_2)}
{\epsilon_1+\epsilon_2- 2\mu}
-
2\,\frac{\theta(\mu-\epsilon_1)\theta(\epsilon_2-\mu)}{\epsilon_1-\epsilon_2}
\biggr\}\,d\epsilon_1\,d\epsilon_2\,\nonumber \ .
\end{eqnarray}
Here $\theta(y)$ is the step function.
The first term in $K(\mu)$  (diagram Fig.~\ref{diag}a) is logarithmically divergent at 
$\epsilon_1,\epsilon_2 \to +\infty$. The divergence is a result of the contact 
approximation (\ref{int}). When the energy is large, $\epsilon \gg 1$, the wave length
is smaller than the size of the barrier, $\lambda \ll 1/\sqrt{m\omega}$, and the contact approximation fails. To fix the problem, we introduce an ultraviolet cutoff $\Lambda$,
$\epsilon_1+\epsilon_2 \le \Lambda$.
Dependence on the cutoff is weak and we will present all results for 
$\Lambda=2$. 
The integrals in (\ref{dg}) cannot be calculated analytically.
However numerical integration is very simple and we present a
plot of $K(\mu)$ in Fig.~\ref{figK}.

\begin{figure}[htb]
\includegraphics[height=100pt,keepaspectratio=true]{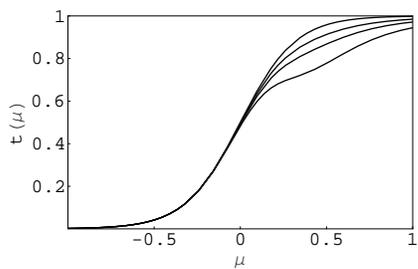}
\vspace{-10pt}
\caption{\it Conductance in units of $2e^2/h$ versus $\mu$ (chemical
potential in units of $\omega$) for different  temperatures:
the intermediate coupling limit, $g_e = 1$.
The uppermost curve corresponds to $T=0$, while the lowest is $T \approx 0.3\textrm{K}$, assuming $\frac{\tau v_F}{L} \approx 1$.} 
\vspace{-15pt}
\label{largeg} 
\end{figure}

Since the coupling constant $g_e \sim 1$, the second order correction alone is not sufficient. However in this regime the Brueckner approximation \cite{br} usually works well.
Since the kernel $K(\mu)$ is independent of external momenta, the Brueckner approximation is
equivalent to the summation of a geometrical progression, and hence the 
renormalized coupling constant $g_R$ is
\begin{equation}
\label{R1}
g_R^2 = g_e^2R(\mu) \ , \ \ \ R(\mu)=\frac{1}{[1-2g_e K(\mu)]^2} \ .
\end{equation}
Plots of $R(\mu)$ for $g_e=1$ and $g_e=2$ are presented in Fig.~\ref{figK}.
Fig.~\ref{largeg} shows a set of plots of conductance ${\overline {t_{\mu}}}$
for different temperatures, using   Eq.~(\ref{dt}) with
$g_e\to g_R(g_e)$  for $g_e=1$.

The results presented in Figs.~\ref{smallg} and \ref{largeg} are very similar
to the experimental data on  the 0.7 structure. According to our calculation 
the 
exact position of the structure depends on the coupling constant $g_e$:
for small $g_e$ it is more like a ``0.5 structure'' and for $g_e\sim 1$ it is
a ``0.6--0.7 structure''.

{\it In conclusion}.
Within perturbation theory, we have considered transport of correlated electrons through 
a quantum point contact.
At zero temperature, the approach results in the usual Landauer formula
and the conductance does not show  any structures. To describe 
the current at nonzero temperature we have formulated a nonlocal kinetic 
equation  for the occupation numbers.
A  current of correlated electrons  scales  as temperature squared
at very low temperatures. 
The  corresponding correction to conductance is negative and strongly enhanced in the
region $0.5\frac{2e^2}{h} \le G \le 1.0\frac{2e^2}{h}$.
We believe that these results are directly relevant to the 0.7 
conductance structure.
For the case of weak coupling, the set of plots of the conductance for 
different temperatures is shown in Fig.~\ref{smallg}. 
The  corresponding plots for  intermediate coupling, which is probably 
the regime most relevant to experiment, is shown in Fig.~\ref{largeg}.
Our model is consistent with the experimental behavior  of the 0.7 structure
under a magnetic field: a field smoothly ``switches off'' the effective
 interaction between electrons.
Effects considered in the present paper have a very simple physical origin:
the electron wave function at the barrier and hence the electron-electron 
interaction is strongly peaked when the transmission coefficient is slightly
higher than 0.5.

\acknowledgments{We are grateful to A.~Hamilton, Y.~Imry, R.~Newbury and M.~Pepper for helpful comments and discussion.
A.~I.~M. gratefully acknowledges the School of Physics at the University of New South Wales 
for warm hospitality and financial support during his visit. 
This work was supported in part by the Australian Research Council.}


\begin{thebibliography}{99}
\bibitem{wharam88} D. A. Wharam {\it et al}, J. Phys. C {\bf 21}, L209 (1988).
\bibitem{vanwees88} B. J. van Wees {\it et al} Phys. Rev. Lett. {\bf 60}, 
848 (1988).
\bibitem{land} R. Landauer, Physics Letters A {\bf 85}, 91 (1981).
\bibitem{buttiker90} M. B\"{u}ttiker, Phys. Rev. B {\bf 41}, 7906 (1990).
\bibitem{thomas96} K. J. Thomas {\it et al}, Phys. Rev. Lett. {\bf 77}, 135 
(1996).
\bibitem{thomas98} K. J. Thomas {\it et al}, Phys. Rev. B {\bf 58}, 
4846 (1998).
\bibitem{kristensen00} A. Kristensen {\it et al}, Phys. Rev. B {\bf 62}, 
10950 (2000).
\bibitem{reilly01} D. J. Reilly {\it et al}, Phys. Rev. B {\bf 63}, 121311 (2001).
\bibitem{cronenwett02} S. M. Cronenwett {\it et al}, Phys. Rev. Lett.
{\bf 88}, 226805 (2002).
\bibitem{noise} P. Roche {\it et al}, Phys. Rev. Lett. {\bf 93}, 116602 (2004). 
\bibitem{danneau} R.~Danneau {\it et al}, Appl. Phys. Lett. {\bf 88}, 012107 
(2006).
\bibitem{Chuan} Chuan-Kui Wang and K.-F. Berggren, Phys. Rev. B {\bf 54},
14257 (1996); Phys. Rev. B {\bf 57}, 4552 (1998).
\bibitem{Calmels} L. Calmels and A. Gold, Solid State Commun. {\bf 106},
139 (1998).
\bibitem{Zabala} N. Zabala, M. J. Puska, and R. M. Nieminen, Phys. Rev. Lett.
{\bf 80}, 3336 (1998).
\bibitem{bruus01} H. Bruus, V. V. Cherepanov, and K. Flensberg, Physica E, 
{\bf 10}, 97 (2001).
\bibitem{starikov03}A. A. Starikov, I. I. Yakimenko, and K.-F. Berggren, 
Phys. Rev. B {\bf 67}, 235319 (2003).
\bibitem{reilly05} D. J. Reilly, Phys. Rev. B {\bf 72}, 033309 (2005).
\bibitem{spivak} B. Spivak and F. Zhou, Phys. Rev. B {\bf 61}, 16730 (2000).
\bibitem{sushkov} O. P. Sushkov, Phys. Rev. B {\bf 64}, 155319 (2001);
{\bf 67}, 195318 (2003).
\bibitem{meir} Y. Meir, K. Hirose, and N. S. Wingreen, Phys. Rev. Lett.
{\bf 89}, 196802 (2002);
K. Hirose, Y. Meir, and N. S. Wingreen, Phys. Rev. Lett.
{\bf 90}, 026804 (2003).
\bibitem{cornaglia04} P. S. Cornaglia and C. A. Balseiro, Europhys. Lett., 
{\bf 67}, 634 (2004).
\bibitem{seelig03} G. Seelig and K. A. Matveev, Phys Rev Lett. {\bf 90}, 176804
(2003).
\bibitem{molina} R.A. Molina {\it et al}, Phys. Rev. B 67, 235306 (2003).
\bibitem{meden} V. Meden and U. Schollwoeck, Phys. Rev. B {\bf 67}, 193303 (2003).
\bibitem{BM} A.~Bohr and B.~Mottelson, {\it Nuclear structure, vol. 2},
Benjamin, New York, Amsterdam, 1974.
\bibitem{buttiker90b} M. B\"{u}ttiker, Phys. Rev. Lett. {\bf 65}, 2901 (1990).
\bibitem{LP} E.~M.~Lifshitz and L~P.~Pitaevskii, {\it Physical kinetics},
Pergamon Press; Oxford , New York; 1981.  
\bibitem{QM} L.~D.~Landau and E.~M.~Lifshitz, {\it Quantum Mechanics},
Pergamon Press;  Oxford, New York; 1965.
\bibitem{br} K.~A.~Brueckner and C.~A.~Levinson, Phys. Rev {\bf 97}, 1344 (1955).
\end{thebibliography}
\end{document}